\documentclass[trackchanges,twocolumn]{aastex7}
\usepackage{hyperref}
\usepackage{mathrsfs}

\begin{document}

\title{Dark gaps and resonances in barred galaxies}

\shorttitle{Dark gaps and resonances}
\shortauthors{Kim et al.}

\author[0000-0002-5857-5136]{Taehyun Kim}
\affiliation{Department of Astronomy and Atmospheric Sciences, Kyungpook National University, Daegu 702-701, Republic of Korea}
\email[show]{tkim.astro@gmail.com}

\author[0000-0003-1775-2367]{Dimitri A. Gadotti}
\affiliation{Centre for Extragalactic Astronomy, Department of Physics, Durham University, South Road, Durham DH1 3LE, UK}
\email{dimitri.a.gadotti@durham.ac.uk}

\author[0000-0003-1544-8556]{Myeong-gu Park}
\affiliation{Department of Astronomy and Atmospheric Sciences, Kyungpook National University, Daegu 41566, Republic of Korea}
\email{mgp@knu.ac.kr}

\author[0000-0003-2779-6793]{Yun Hee Lee}
\affiliation{Department of Astronomy and Atmospheric Sciences, Kyungpook National University, Daegu 41566, Republic of Korea}
\email{yhinjesus@gmail.com}

\author[(0000-0002-0897-3013]{Francesca Fragkoudi}
\affiliation{Institute for Computational Cosmology, Department of Physics, Durham University, South Road, Durham DH1 3LE, UK}
\email{francesca.fragkoudi@durham.ac.uk}

\author[0000-0002-3560-0781]{Minjin Kim}
\affiliation{Department of Astronomy, Yonsei University, 50 Yonsei-ro, Seodaemun-gu, Seoul 03722, Republic of Korea}
\affiliation{Department of Astronomy and Atmospheric Sciences, Kyungpook National University, Daegu 41566, Republic of Korea}
\email{mkim.astro@yonsei.ac.kr}

\author[0000-0003-4625-229X]{Woong-Tae Kim}
\affiliation{Department of Physics \& Astronomy, Seoul National University, Seoul 08826, Republic of Korea}
\affiliation{SNU Astronomy Research Center, Seoul National University, Seoul 08826, Republic of Korea}
\email{wkim@astro.snu.ac.kr}

\begin{abstract}
Dark gaps, low surface brightness regions along the bar minor axis, are expected to form as a consequence of secular evolution in barred galaxies. Although several studies have proposed links between dark gap locations and dynamical resonances, the results remain inconclusive. Using DESI Legacy Imaging Survey data, we find that approximately 61$\%$ of barred galaxies exhibit pronounced dark gaps. We compare the location of dark gaps with resonance radii derived from the Tremaine–Weinberg method applied to MaNGA data for the same galaxies. Our analysis shows that dark gaps do not preferentially form at specific resonances. Instead, their locations correlate with $\mathcal{R} \equiv R_{\rm CR}/R_{\rm Bar}$: slow bars tend to show shorter dark gap radii, while fast bars show longer ones. This trend reflects a tight relation between bar length and dark gap radius. However, when barred galaxies are classified by their ring morphology, certain types exhibit dark gaps that align with specific resonances. Notably, dark gaps located between the inner and outer rings are closely associated with the corotation radius. In galaxies with two dark gaps along the bar minor axis profile, the inner dark gap typically aligns with the ultraharmonic resonance, and the outer dark gap corresponds to the corotation radius.
These findings suggest that some morphological types share similar $\mathcal{R}$ values and exhibit dark gaps near specific resonances. Thus, dark gaps may serve as proxies for dynamical resonances only in certain systems. Our findings may help explain the discrepancies observed in earlier studies.
\end{abstract}



\section{Introduction} 
Stellar bars are commonly found in nearby disk galaxies (e.g., \citealt{eskridge_00, menendez_delmestre_07, sheth_08, aguerri_09, barazza_09, marinova_09, masters_11, buta_15, lee_19, wang_25}).
Numerical simulations suggest that, as barred galaxies evolve, bars typically increase in length and strength while their pattern speeds gradually decrease, primarily due to angular momentum transfer to the dark matter halo and bulge (e.g., \citealt{weinberg_85, debattista_00, athanassoula_03, athanassoula_13, fragkoudi_21, jang_23, jang_24}).
In particular, theoretical models suggest that as galaxies evolve, bars trap stars on quasi-circular orbits just outside the bar, which are subsequently captured into the elongated $x_1$ orbits that support the bar structure or higher order resonances \citep{contopoulos_89b, athanassoula_03, binney_08}. This process leads to bars becoming longer, more massive, and dynamically stronger over time \citep{athanassoula_03}.
As stars are extracted from the inner disk and redistributed into the bar, the surrounding inner disk region becomes depleted, creating low-density zones \citep{gadotti_03}. These regions are often referred to as dark gaps \citep{buta_17b} or described as a ``light deficit around the bar'' \citep{kim_16}. They have also been characterized as ``star formation deserts" due to their lack of ongoing star formation \citep{james_09, james_18, donohoe_keyes_19}. 
\textit{N}-body simulations suggest that mass loss along the bar minor axis can reach up to 60 –- 80$\%$ of the initial mass in these regions \citep{ghosh_24b}. The associated mass deficits in these dark gaps tend to be more prominent in galaxies hosting longer and stronger bars \citep{kim_16, kim_21, aguerri_23, ghosh_24b}.

\citet{buta_17b} proposes that the dark gaps observed between the inner and outer rings of barred galaxies are likely associated with the locations of the $L_4$ and $L_5$ Lagrangian points. As these points are theoretically expected to lie near the corotation resonance (CR) of the bar, the author suggests that the positions of dark gaps can be used to infer the CR radius of the bar. Earlier numerical studies support this interpretation which find that the regions near the $L_4$ and $L_5$ points are typically depopulated in barred galaxies \citep{Pfenniger_90a, schwarz_81, schwarz_84c, byrd_94c, salo_99}. These depopulated zones may correspond to the observed dark gaps in the stellar light distribution.

Other studies find that dark gaps are associated with resonances other than corotation. \citet{krishnarao_22}, using the high-resolution GALAKOS N-body simulation \citep{donghia_20}, find that dark gaps are more closely associated with the inner ultraharmonic resonance (UHR). 
It is worth noting that the simulated galaxy in their study does not develop a ring — one of the key assumption in \citet{buta_17b} that associates dark gaps with corotation. 
\citet{aguerri_23} analyzed 37 barred galaxies with pattern speeds measured via the Tremaine–Weinberg (TW) method (\citealt{tremaine_84a}) and find that most dark gaps lie near the UHR, though a significant minority (10 out of 37) are located close to corotation. 


On the other hand, \citet{ghosh_24b} examine a set of N-body simulations that include both thin and thick disk components and find that the locations of dark gaps do not align with any of the classical resonances (CR, UHR, or the inner Lindblad resonance).

As outlined above, although various interpretations have been proposed to explain the connection between dark gaps and galactic resonances, significant discrepancies remain and no clear consensus has yet been established. This inconsistency highlights the need for further observational studies to elucidate the physical origin of dark gaps and their connection to bar dynamics.


Barred galaxies evolve through the redistribution of angular momentum (e.g., \citealt{lyndenbell_72, debattista_00, athanassoula_03, martinez_valpuesta_06, sellwood_14_rev}). 
This exchange of angular momentum occurs mainly at dynamical resonances associated with the bar or spiral structure. In the inner disk, however, bar-driven resonances dominate the dynamics.
The bulk of angular momentum in the inner disk is lost near the inner Lindblad resonance (ILR) and subsequently absorbed by the dark matter halo, stellar bulge, and outer disk, primarily at the CR, though other resonances also contribute to a lesser extent \citep{athanassoula_03, weinberg_07b, dubinski_09, saha_13b, donghia_20, jang_23, li_25xx, mcclure_25b_xx}.
Consequently, identifying and characterizing the radii of key resonances is essential for understanding the mechanisms of angular momentum transfer that drive the secular evolution of barred galaxies.


In most studies, resonance radii are estimated under the assumption of the epicyclic approximation, which is valid only in the case of weak perturbations. Therefore, caution is required when applying this framework to galaxies with strong bars. In particular, it is likely more accurate to describe resonances as extended regions rather than radii (see \citealt{contopoulos_80a}).
A straightforward method for estimating resonance radii is to measure the bar pattern speed ($\Omega_{\rm bar}$), which represents the angular rotation rate of the bar pattern. $\Omega_{\rm bar}$ is a fundamental parameter that characterizes the dynamical state of barred galaxies. 
The corotation radius ($R_{\rm CR}$) is the radius at which the angular rotation speed of the disk material matches $\Omega_{\rm bar}$.
A key dimensionless parameter in bar dynamics is the ratio of the corotation radius to the bar length, defined as $\mathcal{R} \equiv R_{\rm CR} / R_{\rm bar}$. 
This ratio is widely employed to classify bars as either fast ($\mathcal{R} \leq 1.4$) or slow ($\mathcal{R} > 1.4$) bars \citep{debattista_00}.
Numerical simulations predict that, as galaxies evolve, bars gradually slow down due to dynamical friction with the surrounding dark matter haloes \citep{algorry_17, peschken_19, roshan_21b}. Therefore $\mathcal{R}$ has been widely used as a diagnostic to test this theoretical expectation. However, observations find that bars predominantly rotate fast \citep{aguerri_15, cuomo_19b, guo_19}, creating a tension between theoretical predictions and observational results.
However, \citet{fragkoudi_21}, using the AURIGA simulations, show that fast bars can arise in more baryon-dominated systems, offering a potential resolution to this tension.

\citet{tremaine_84a} introduced a kinematic method to measure $\Omega_{\rm bar}$, which has since become widely known as the TW method and remains the most widely used approach.
The TW method is straightforward and does not rely on any dynamical modeling. It assumes that the disk possesses a single well-defined $\Omega_{\rm bar}$, the disk is flat and in a steady state, and the tracer population obeys the continuity equation. 
To satisfy the continuity equation, the TW method was initially applied to non-star-forming SB0 galaxies (e.g., \citealt{kent_87, debattista_02a, aguerri_03}). However, with the use of stellar kinematics, the method has since been extended to later Hubble types (e.g., \citealt{gerssen_07, aguerri_15, guo_19, garma-oehmichen_20, williams_21}). Nonetheless, applying the TW method to actively star-forming galaxies requires caution, as the continuity equation may not strictly hold.

Alternative methods have been developed, including: comparison with gas dynamical simulations \citep{weiner_01b, salo_99, rautiainen_05, treuthardt_12, sormani_15b, fragkoudi_17a}; 
phase-shift analysis in the color profiles of galaxies \citep{beckman_90, puerari_97, sierra_15}; 
phase-shift between the gravitational potential and the density based on dynamical considerations \citep{zhang_07, buta_09b}; 
offsets between gas and star formation regions \citep{egusa_09}; 
phase reversals in the velocity field \citep{font_11b, font_14a, beckman_18}; gravitational torque mapping \citep{garcia_burillo_05, ruiz-garcia_24}; 
and morphological indicators \citep{buta_17b}.

In this study, we aim to examine whether dark gaps are physically associated with specific resonances by analyzing observational datasets spanning a diverse sample of barred galaxies. The structure of the paper is as follows. Section 2 describes the dataset and analysis methodology. Section 3 explores the locations and frequencies of dark gaps using two estimation methods. In Section 4, we compare the estimated positions of dark gaps with corotation radii (CR) derived from the TW method. Section 5 evaluates whether specific morphological types exhibit a preferred association between dark gaps and resonances. Section 6 discusses the origin of previously reported discrepancies between resonance radii and observed dark gap locations. Finally, Section 7 presents the summary and conclusions.


\section{Sample \& Data} 
\subsection{Sample with kinematics}
We selected a sample of barred galaxies for which $\Omega_{\mathrm{bar}}$ have been consistently measured. Specifically, we used the sample from \citet{geron_23}, who determined the quantities for 225 barred galaxies using integral field unit data from the Mapping Nearby Galaxies at APO (MaNGA) survey \citep{bundy_15}. The bar pattern speeds were derived using the TW method \citep{tremaine_84a}. To date, this sample represents the largest collection of barred galaxies with consistently measured $\Omega_{\mathrm{bar}}$ and $R_{\rm CR}$. It spans a wide range of stellar masses ($10 \leq \log (M_{*}/M_{\odot}) \leq 11.5$), morphologies, and bar strengths, including weak and strong bars.

Barred galaxies were selected from the Galaxy Zoo DESI catalog \citep{Walmsley_23}. Galaxy classifications were provided by deep learning models trained on volunteer responses from the Galaxy Zoo DECaLS campaign \citep{walmsley_22}. The final models were trained using a dataset of 401,000 galaxies with a total of 10 million volunteer classifications.

To meet the requirements of the TW method, galaxies are selected to have inclinations between $20^\circ$ and $70^\circ$, and a minimum angular separation of $10^\circ$ between the bar and the major or minor axis of the galaxy. 
A detailed description of the sample selection and the methodology used to derive the bar pattern speeds is provided in \citet{geron_23}. 
In order to obtain $R_{\rm CR}$, \citet{geron_23} derived the rotation curve of the galaxy from stellar velocity measurements, using a 5 arcsec aperture aligned with the major axis of the galaxy. They then determined $R_{\rm CR}$ as the radius at which the rotation curve intersects with the line defined by $\Omega_{bar} \times R$. For further details, see Section 2.3 of  \citet{geron_23}.


\begin{figure*}[ht!]
    \centering
    \includegraphics[width=0.495\textwidth]{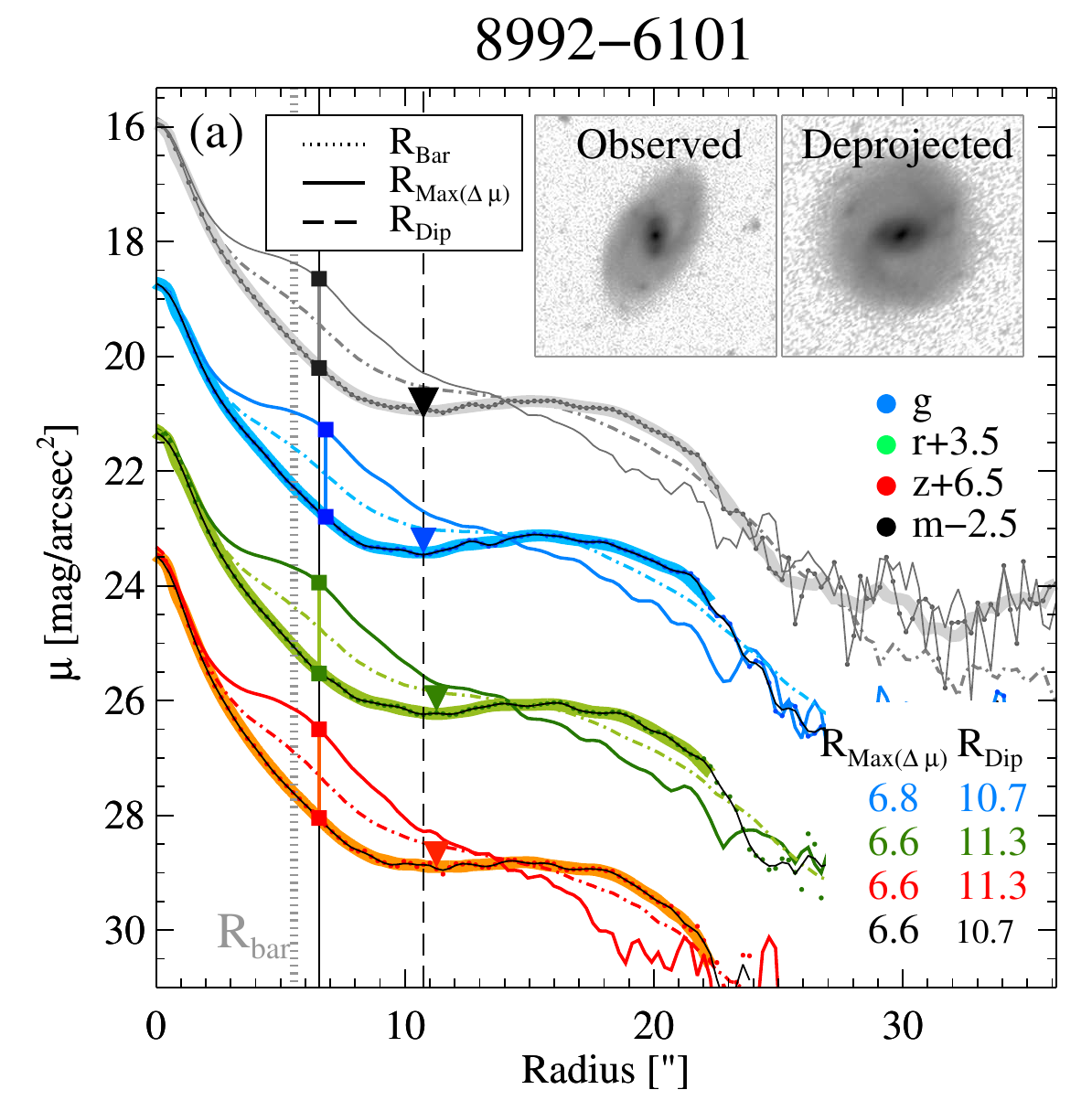}
    \includegraphics[width=0.495\textwidth]{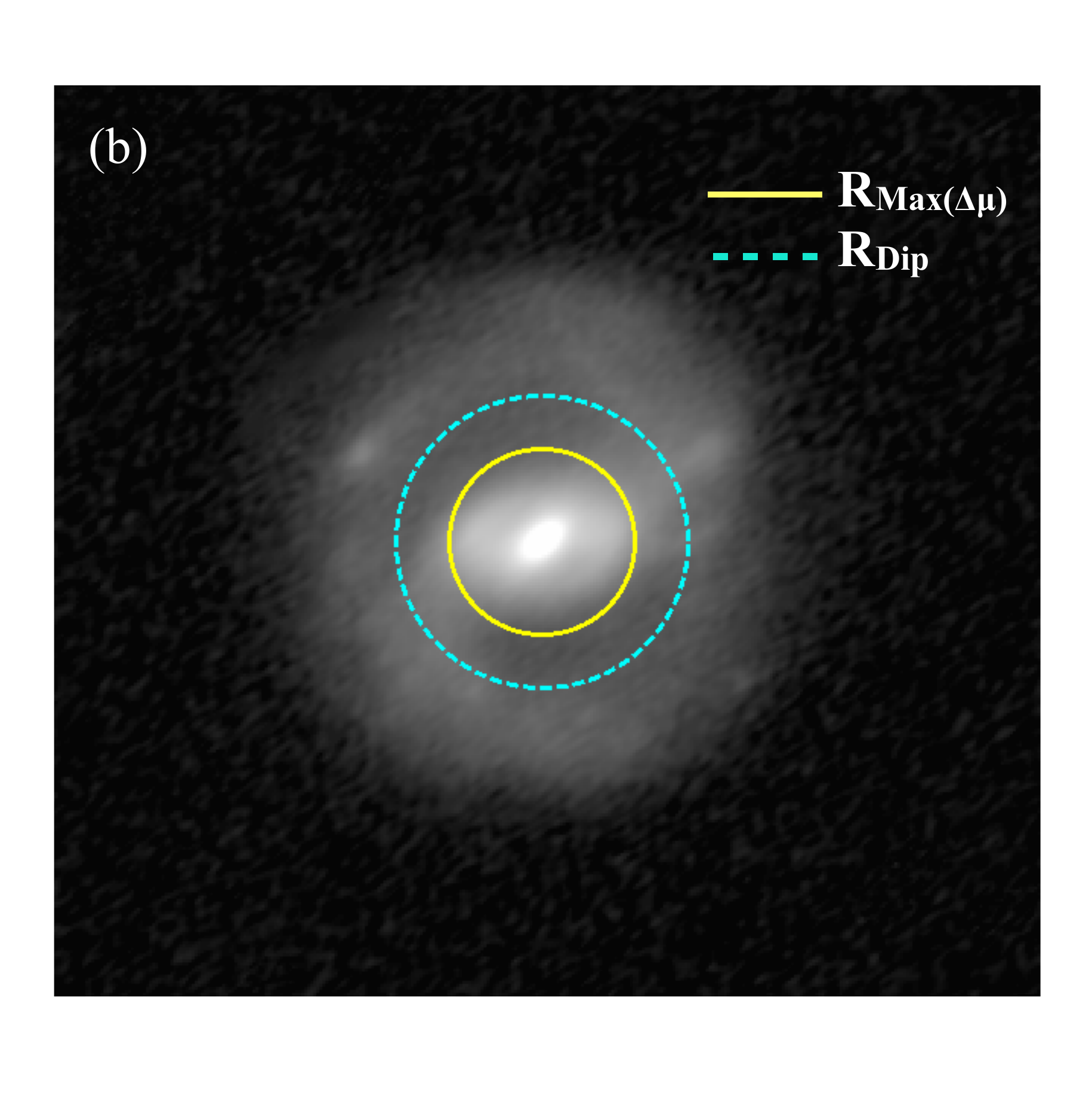}
\caption{
(a): Radial surface brightness profiles of the deprojected galaxy MaNGA 8992-6101 are shown for the $g$ (blue), $r$ (green) and $z$ (red) bands. The average of all available bands at each radius is plotted in black and labeled as “m” (master) in the plot. A 3-pixel-wide pseudo-slit is placed along the bar major and minor axes to obtain the mean surface brightness at each radius. The radial profile along the bar major axis is plotted as a thin solid line, while the profile along the bar minor axis is shown as a thick solid line with markers at each radial point. Azimuthally averaged radial profiles are plotted as dash-dotted lines. Vertical line segments with square caps on both ends indicate the radius corresponding to the maximum difference in surface brightness profiles along the bar major and minor axes, $R_{\mathrm{Max}(\Delta \mu)}$.
Triangles indicate the dip radius ($R_{\mathrm{Dip}}$) in the radial profile along the bar minor axis for each band. Vertical lines represent the bar radius ($R_{\mathrm{bar}}$, shown as a dotted line), the radius of the maximum difference between radial profiles along the bar major and minor axes ($R_{\mathrm{Max}(\Delta \mu)}$, solid line), and the dip radius along the bar minor axis ($R_{\mathrm{Dip}}$, dashed line). The values of $R_{\mathrm{Max}(\Delta \mu)}$ and $R_{\mathrm{Dip}}$ are displayed at the bottom right of the plot, where the measurements from the “m” (master) profile are used for further analysis.
(b): $g$-band image of MaNGA 8992-6101. The yellow solid circle marks $R_{\rm Max(\Delta \mu)}$, while the cyan dashed circle denotes $R_{\rm Dip}$. The black solid bar in the bottom-left corner represents $10''$.
\label{fig:fig1_radial}
}
\end{figure*}

\subsection{Photometric data: DESI} \label{subsec:data}

We utilize optical imaging data from the Dark Energy Spectroscopic Instrument (DESI) Legacy Imaging Surveys Data Release 10 (DR10; \citealt{dey_19}). 
We use these images to investigate dark gaps of galaxies in our sample. The DESI Legacy Surveys provide $g$, $r$, and $z$-band images obtained from three major imaging surveys: the Dark Energy Camera Legacy Survey (DECaLS; \citealt{flaugher_15}), the Beijing–Arizona Sky Survey (BASS; \citealt{zou_17_bass}), and the Mayall $z$-band Legacy Survey (MzLS; \citealt{dey_16, dey_19}). As a result, all galaxies in our sample are covered in the $g$, $r$, and $z$ bands. In addition, DR10 includes $i$-band observations from a variety of non-DECaLS surveys conducted with the Dark Energy Camera (DECam), including the Dark Energy Survey, DELVE, and the DeROSITA Survey. Through this extended coverage, $i$-band images are also available for 75 galaxies in our sample.

The images have a pixel scale of $0.262''$/pixel, and the median full width at half maximum (FWHM) of the $r-$band DESI images ranges from $1.2''$ to $1.5''$. The median 5$\sigma$ detection limit is 23.3 AB mag in the $r-$band for DECaLS, 22.9 AB mag for BASS, and 22.3 AB mag in the $z-$band for MzLS (see Table 4 of \citealt{dey_19}).

To obtain an accurate and physically meaningful deprojection of the image, we applied additional selection steps to the initial sample of 225 barred galaxies identified by \citet{geron_23}.
Using the inclination and the PA of the galaxy and the bar provided in their catalog, we deprojected each galaxy image and rotated it so that the bar major axis is aligned horizontally. 
Then, we visually inspect these images to ensure that the outer isophotes appear circular and that the bars are properly aligned. This facilitates the extraction of surface brightness profiles along both the major and minor axes of the bar. 
Galaxies are further required to exhibit a clearly identifiable bar structure. After applying these selection steps, our final sample consists of 193 galaxies. Among these galaxies, $R_{\rm CR}$ has been estimated for 179 barred galaxies in \citet{geron_23}. 


\begin{figure*}[ht!]
    \centering
    \includegraphics[width=0.325\textwidth]{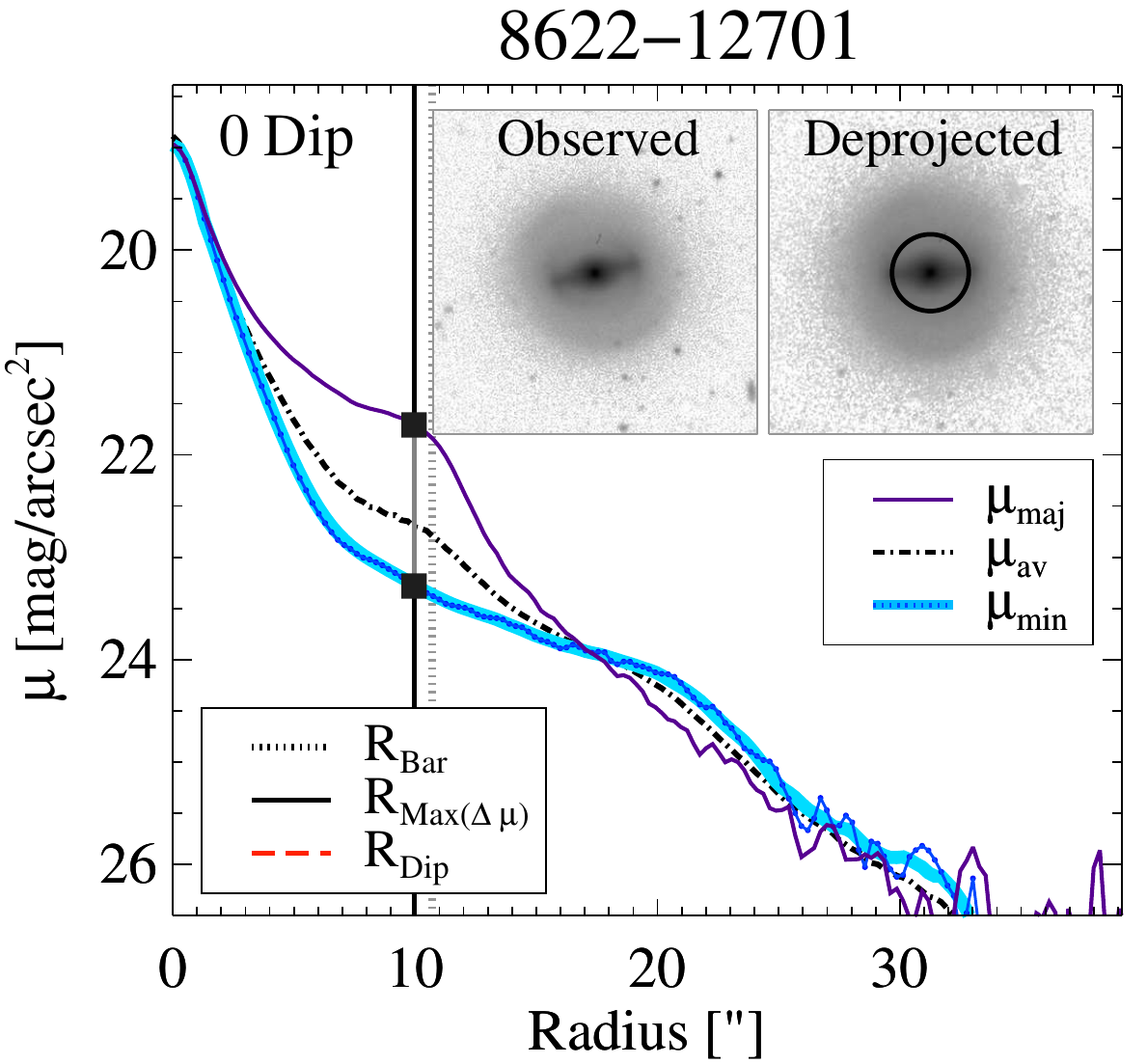}
    \includegraphics[width=0.325\textwidth]{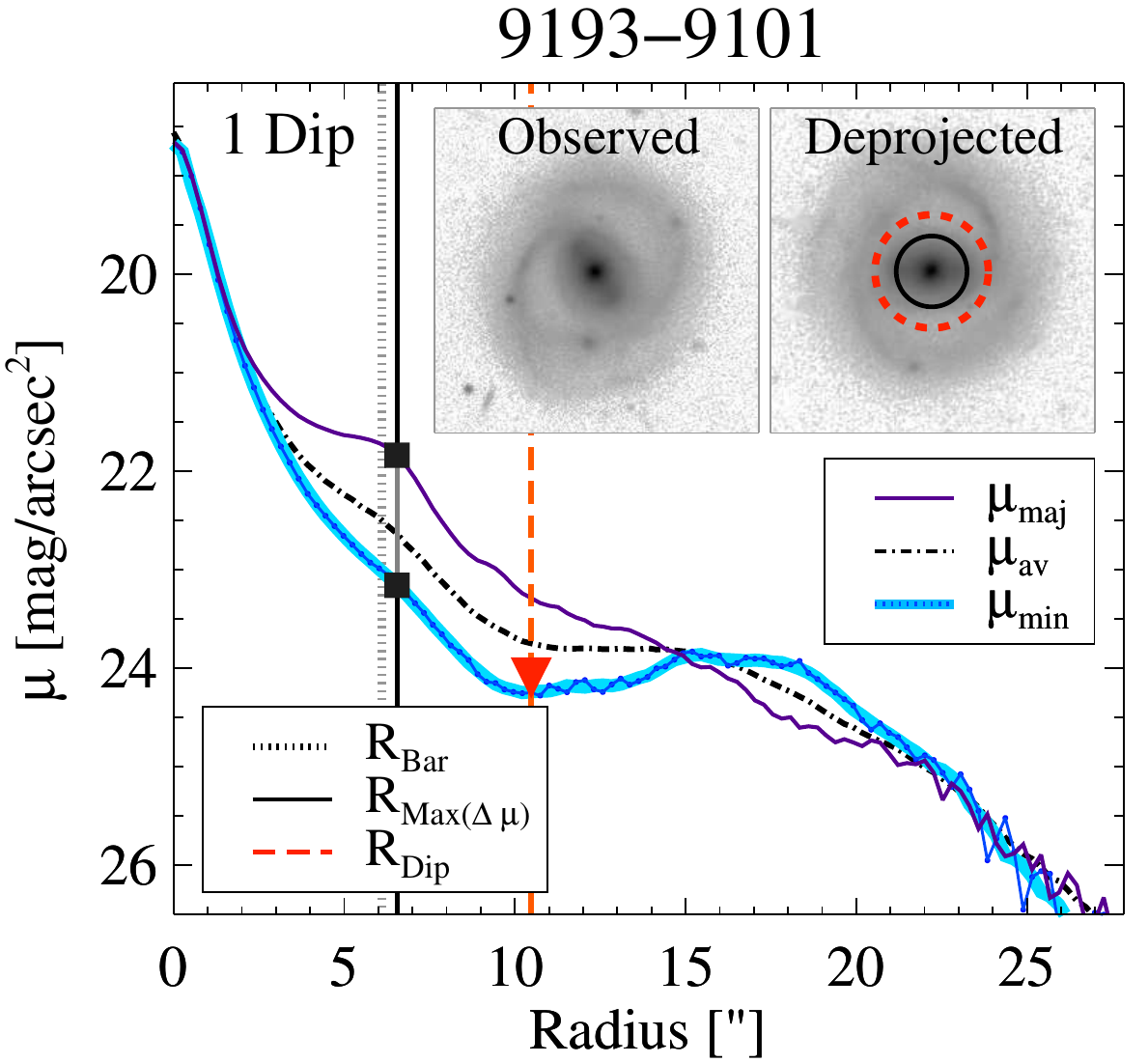}
    \includegraphics[width=0.325\textwidth]{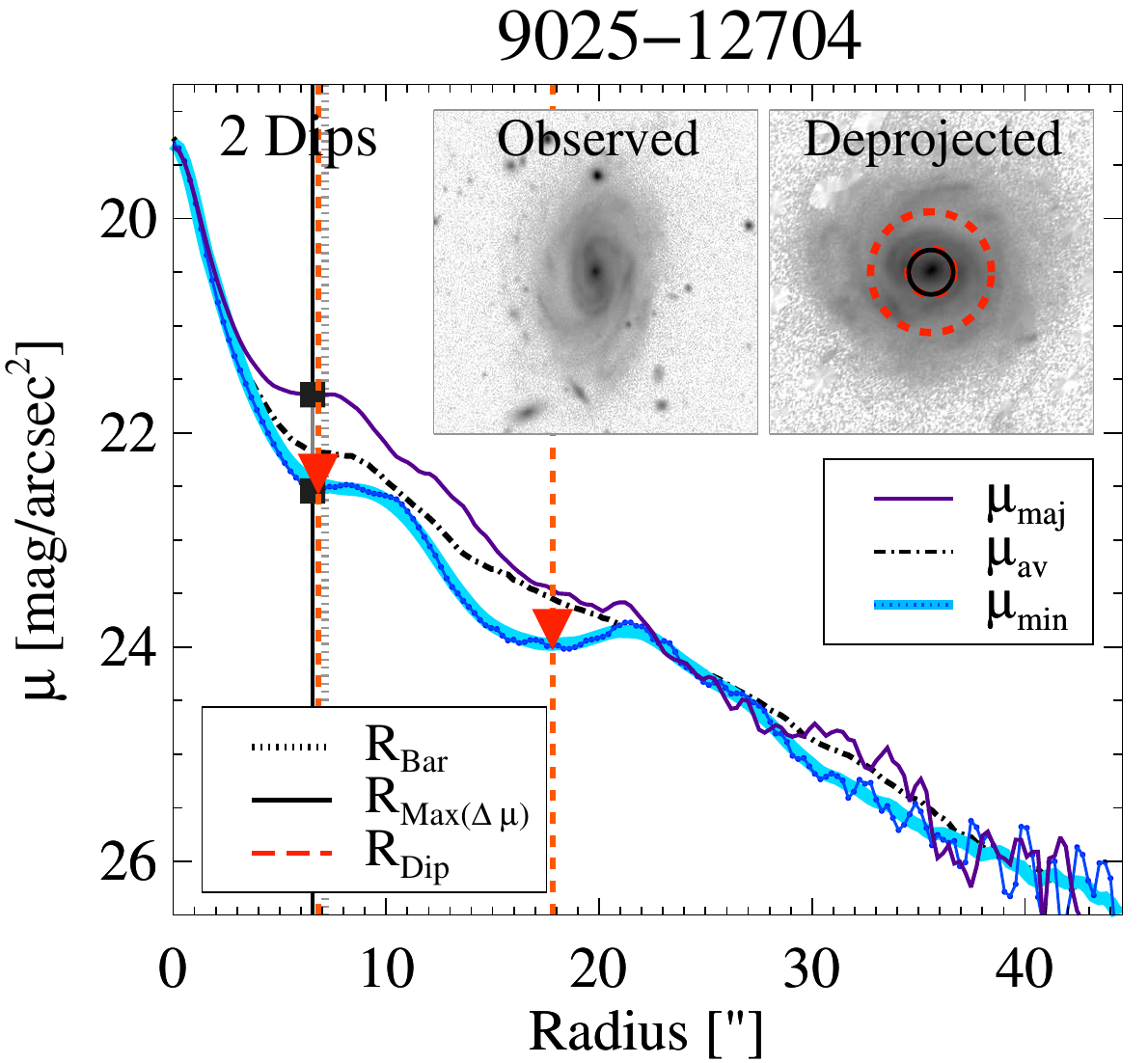}
    \caption{Radial surface brightness profiles of galaxies, derived by averaging data over all available $g$, $r$, $i$ (if present), and $z$ bands, are shown. Three types of barred galaxies are presented, categorized by the presence of dips along the bar minor axis: (1) no dip, (2) a single dip, and (3) two dips. Thin solid lines represent the radial profiles along the bar major axis, while thick blue lines denote the profiles along the bar minor axis. Azimuthally averaged profiles are indicated by dot-dashed lines. Dips are marked with red triangles in the profile and the $R_{\rm Dip}$ is shown in orange dashed circle in the deprojected image, if any. The maximum differences between the bar major and minor axis profiles are shown as lines with black square caps in the profile and $R_{\mathrm{Max}(\Delta \mu)}$ is shown in black circle in the deprojected image.}
    \label{fig:three_in_row}
\end{figure*}

\begin{figure*}[htb!]
\includegraphics[width=0.95\textwidth]{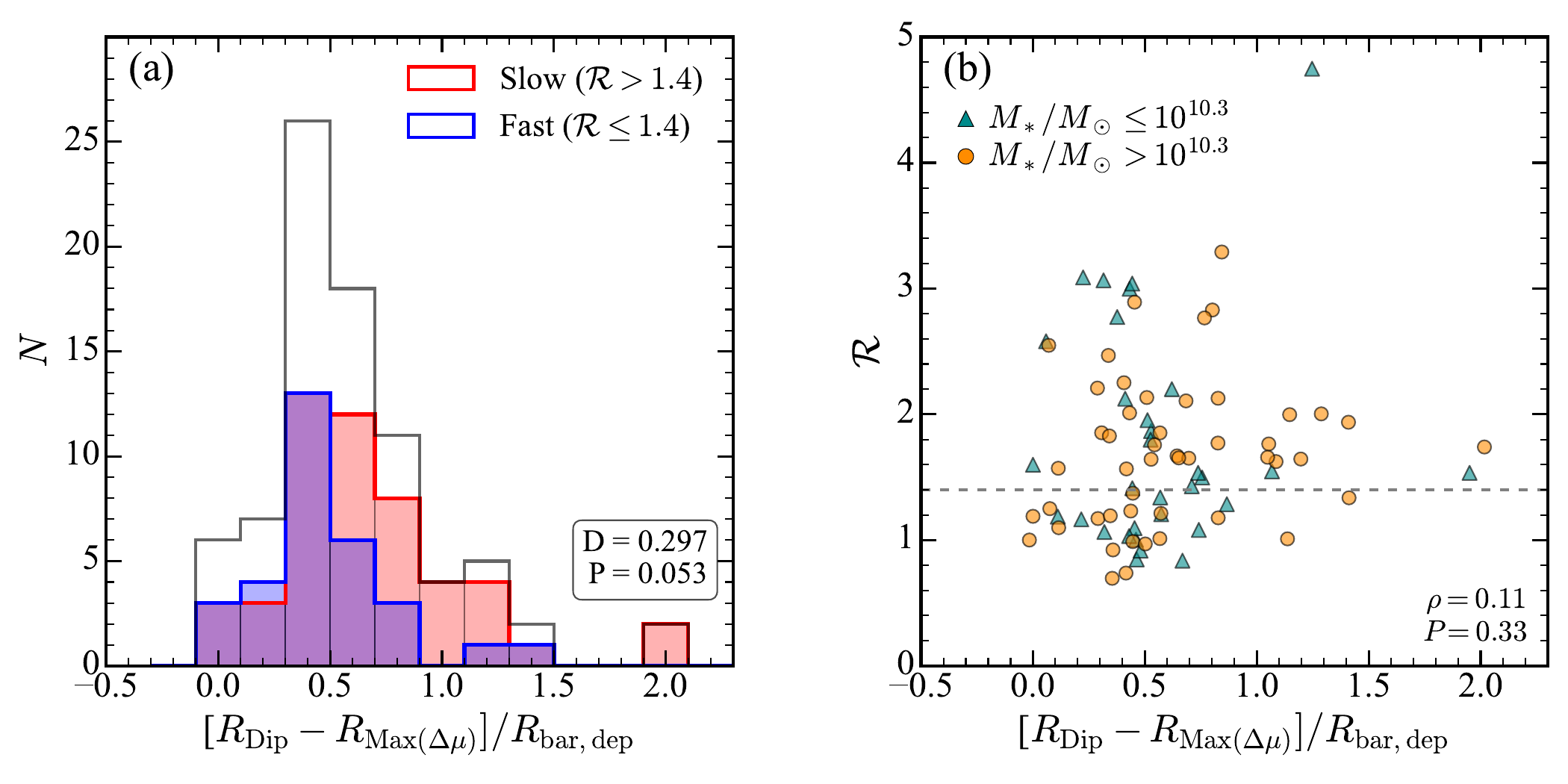}
\caption{(a) Histogram of the difference between $R_{\mathrm{Max}(\Delta \mu)}$ and $R_{\mathrm{Dip}}$, normalized by the deprojected bar length, for galaxies exhibiting a single dip. The sample is divided into fast and slow bars.
(b) Normalized difference between the two dark gap radii and the bar rotation rate, $\mathcal{R}$, which is $R_{\rm CR}/R_{\rm bar}$. Horizontal dashed line divides fast ($\mathcal{R}$ $\leq$ 1.4) and slow ($\mathcal{R}$ $>$ 1.4) bars. Data points are color-coded by total stellar mass.}
\label{fig:fig3_diff}
\end{figure*}

\section{Locations and fractions of dark gaps in galaxies} \label{bsec:data_analysis}

To determine the precise locations of dark gaps, we perform the following analysis on the galaxy images. Background galaxies and foreground stars are masked using source detections obtained with {\sc SExtractor} \citep{bertin_96}. 
Surface brightness is measured along these axes using 3-pixel-width pseudo slits, and the brightness values are averaged at each radius.
Figure~\ref{fig:fig1_radial}(a) shows the radial surface brightness profiles of MaNGA 8992-6101, a barred galaxy with an outer ring, located at z=0.034 \citep{Abazajian_04}. 
The radius of the dark gap is estimated with two methods. First, we measure the location where the surface brightness difference between the bar major and minor axes reaches its maximum \citep{kim_16, krishnarao_22, aguerri_23, ghosh_24b}, hereafter denoted as $R_{\rm Max(\Delta \mu)}$. This radius corresponds to the point where the surface density difference induced by the bar - between the major and minor axes - reaches its maximum, making it a useful diagnostic of bar-driven structural asymmetries.
Second, we find that there is a dip (local minimum) in the surface brightness profile along the bar minor axis at a radius denoted by $R_{\rm Dip}$, located beyond $R_{\rm Max(\Delta \mu)}$. This dip is not evident in the major-axis profile or in the azimuthally averaged radial profile, indicating that it is a localized feature best detected along the bar minor axis, likely associated with the dark gap. A similar method was used to measure the dark gap radius by \citet{buta_17b}.
To enhance the signal-to-noise ratio and mitigate band-specific variations, we construct a master radial profile (hereafter referred to as the “$m$”-band) by averaging the surface brightness profiles from the available $g$, $r$, $i$ (if available), and $z$ bands at each radius. Both $R_{\rm Max(\Delta \mu)}$ and $R_{\rm Dip}$ are measured from this combined profile and are used in all subsequent analyses.
Figure~\ref{fig:fig1_radial}(b) presents the spatial locations of $R_{\rm Max(\Delta \mu)}$ and $R_{\rm Dip}$ overlaid on the $g$-band image of the sample galaxy.

The sample is classified into three categories based on radial surface brightness profiles measured along the bar minor axis: (i) galaxies exhibiting no dip, (ii) galaxies with a single dip, and (iii) galaxies with two dips. Representative examples are presented in Figure~\ref{fig:three_in_row}.
It is possible that the different types of dips are related to the presence of rings, lenses, and spirals. We will further discuss this in Section 6.4.
We find that 45.6$\%$ (88/193) of the barred galaxies exhibit a single dip, while 15.5$\%$ (30/193) show two dips. In total, 61.1$\%$ (123/193) of the sample display at least one dip along the bar minor axis. The remaining 38.9$\%$ (75/193) do not show any such feature, but showing monotonically decreasing profile lacking any local extrema. If we define a galaxy as hosting a pronounced dark gap when it shows at least one dip along the bar minor axis, our result indicates that the majority of barred galaxies in our sample exhibit pronounced dark gaps. 

The fraction of galaxies with dark gaps in our sample is notably higher than the 32$\%$ reported by \citet{krishnarao_22}, who used stellar mass surface density maps from MaNGA datasets.
Several factors may contribute to this discrepancy, which will be discussed in more detail in Sec.~\ref{subsec:no_dg}. One possible reason is that the DESI multiband images are deeper than the MaNGA dataset, allowing us to identify more barred galaxies exhibiting dark gaps. In this study, we examine radial profiles using combined images from the $g$, $r$, $i$ (if available), and $z$ bands, whereas \citet{krishnarao_22} base their analysis on stellar mass maps derived from MaNGA via principal component analysis (PCA). 
The discrepancy may partly result from differences in image resolution, as dark gaps observed in our high-resolution data could be smoothed out in lower-resolution images.
Additionally, the criteria used to define dark gaps may differ between the two studies. Here, we classify a galaxy as exhibiting a clear dark gap if a dip is present in the radial profile along the bar minor axis. By contrast, \citet{krishnarao_22} do not provide an explicit definition of how dark gaps are identified, making a direct comparison between the two results difficult.

Even galaxies that do not exhibit a dip clearly show enhanced brightness along the major axis of the bar compared to the minor axis, i.e., a noticeable $\rm Max(\Delta \mu)$. Although barred galaxies are classified based on the presence or absence of dips, we note that it is the same physical process that is responsible for both cases — namely, the trapping of stars by the bar.

Figure~\ref{fig:fig1_radial} clearly demonstrates that $R_{\rm Max(\Delta \mu)}$ and $R_{\rm Dip}$ are distinct.
To quantify this difference, we present in Figure~\ref{fig:fig3_diff} the distribution of their separation, normalized by the deprojected bar radius.
We find that, for the majority of barred galaxies with a single dip, the two radii differ significantly: in 80.7$\%$ (71/88) of such galaxies, the difference exceeds 0.3$\times R_{\rm bar}$, and in more than 50.0$\%$ (44/88), it exceeds 0.5$\times R_{\rm bar}$.
In most galaxies with a single dip, $R_{\rm Max(\Delta \mu)}$ is located near the end of the bar, where the radial profile along the bar major axis exhibits a ``shoulder''— an excess in surface density above the exponential disk profile (\citealt{anderson_22, beraldo_e_silva_23, erwin_23}). In contrast, $R_{\rm Dip}$ is typically found at larger radii. This spatial offset results in a systematic difference between $R_{\rm Max(\Delta \mu)}$ and $R_{\rm Dip}$.

In Figure~\ref{fig:fig3_diff}, we also present the distribution of the differences between $R_{\rm Max(\Delta \mu)}$ and $R_{\rm Dip}$ for fast bars in blue and slow bars in red. A Kolmogorov–Smirnov (KS) test yields a statistic of $D = 0.297$ and a $p$-value of 0.054 between the two populations.
This suggests a marginal difference between the distributions that does not reach conventional levels of statistical significance ($p < 0.05$). To further investigate this, we plot the $\mathcal{R}$ and normalized difference between the two dark gap radii in Figure~\ref{fig:fig3_diff}(b), defined as $(R_{\rm Dip} - R_{\rm Max(\Delta \mu)}) / R_{\rm bar, dep}$.
Pearson correlation coefficients indicate that there is no significant correlation between this normalized difference and $\mathcal{R}$. Similarly, we find no clear dependence on stellar mass.

Interestingly, for barred galaxies exhibiting two dips, the first dip radius ($R_{\rm 1st/2dips}$) generally agrees well with $R_{\rm Max(\Delta \mu)}$, as shown in Figure~\ref{fig:three_in_row} for galaxies with two dips.
Among the 30 galaxies with two dips, only one shows a difference between $R_{\rm 1st/2dips}$ and $R_{\rm Max(\Delta \mu)}$ that exceeds 0.3$\times R_{\rm bar}$.
In contrast, the radius of the second dip ($R_{\rm 2nd/2dips}$) differs from $R_{\rm Max(\Delta \mu)}$ by more than 0.3$\times R_{\rm bar}$ in 93$\%$ (28/30) of galaxies with two dips.
Therefore, the distinction between the two radii should be carefully considered when analyzing dark gaps.

\begin{figure*}[ht!]
\includegraphics[width=\textwidth]{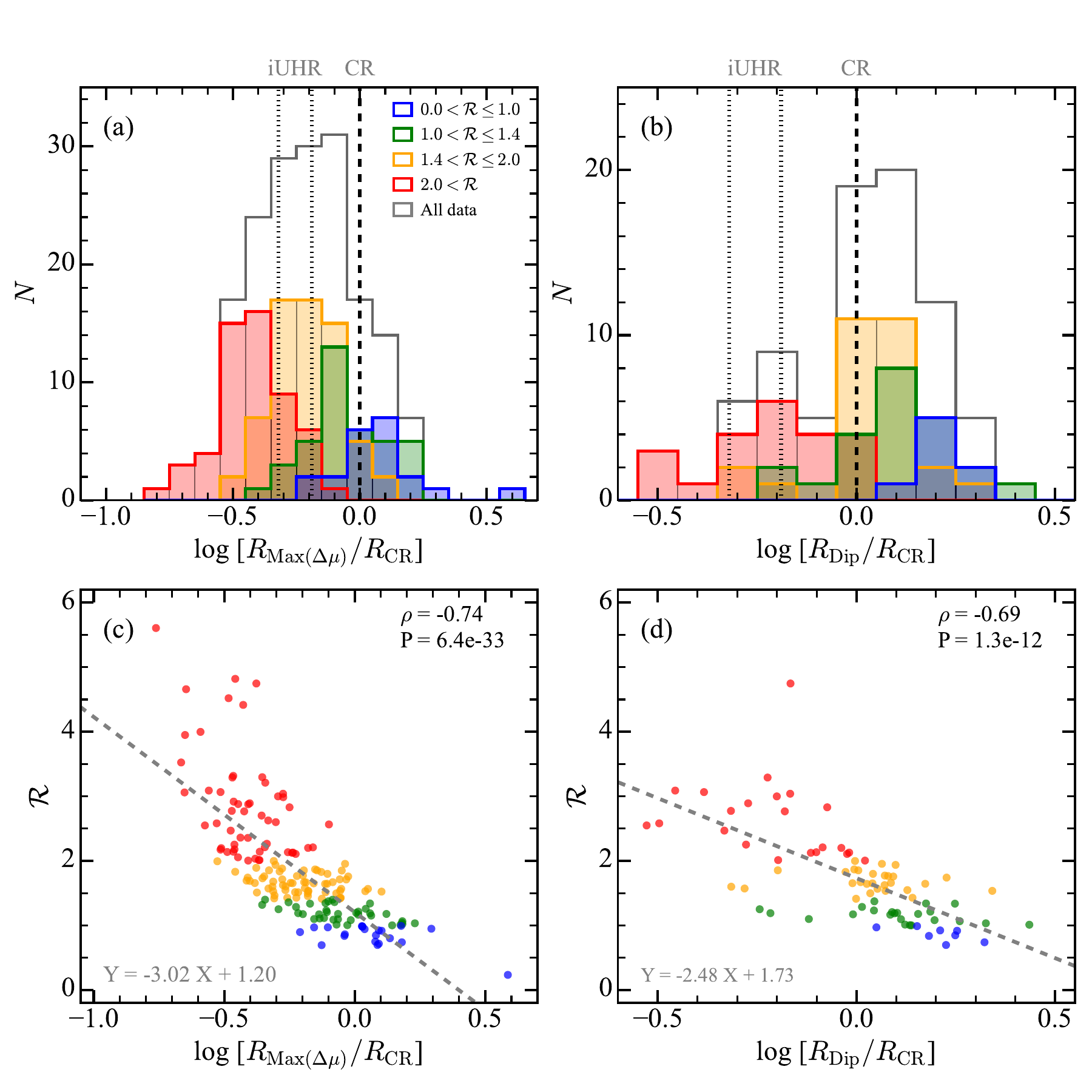}
\caption{
Ratio of dark gap radius to corotation radius.
(a): Histogram of $\log(R_{\mathrm{Max}(\Delta \mu)}/R_{\mathrm{CR}})$ for all galaxies. The vertical dashed line indicates where $R_{\mathrm{Max}(\Delta \mu)}$ equals $R_{\mathrm{CR}}$, while the region enclosed by the two dotted lines represents the inferred extent of the inner UHR, estimated using Equation~\ref{eq:resonances}.
(b): Histogram of $\log(R_{\mathrm{Dip}}/R_{\mathrm{CR}})$ for galaxies exhibiting a single dip. 
(c): Relation between $\log(R_{\mathrm{Max}(\Delta \mu)}/R_{\mathrm{CR}})$ and the rotation rate, $\mathcal{R}$. Spearman’s rank correlation coefficient ($\rho$) and statistical significance (P) are shown in the upper right. A linear fit to the data is displayed as a gray dashed line, with the corresponding equation given in the lower left corner. 
(d): Relation between $\log(R_{\mathrm{Dip}}/R_{\mathrm{CR}})$ and $\mathcal{R}$ for galaxies with a single dip.
}
\label{fig:rdg_rcr}
\end{figure*}

\begin{figure}[htb!]
\includegraphics[width=0.49\textwidth]{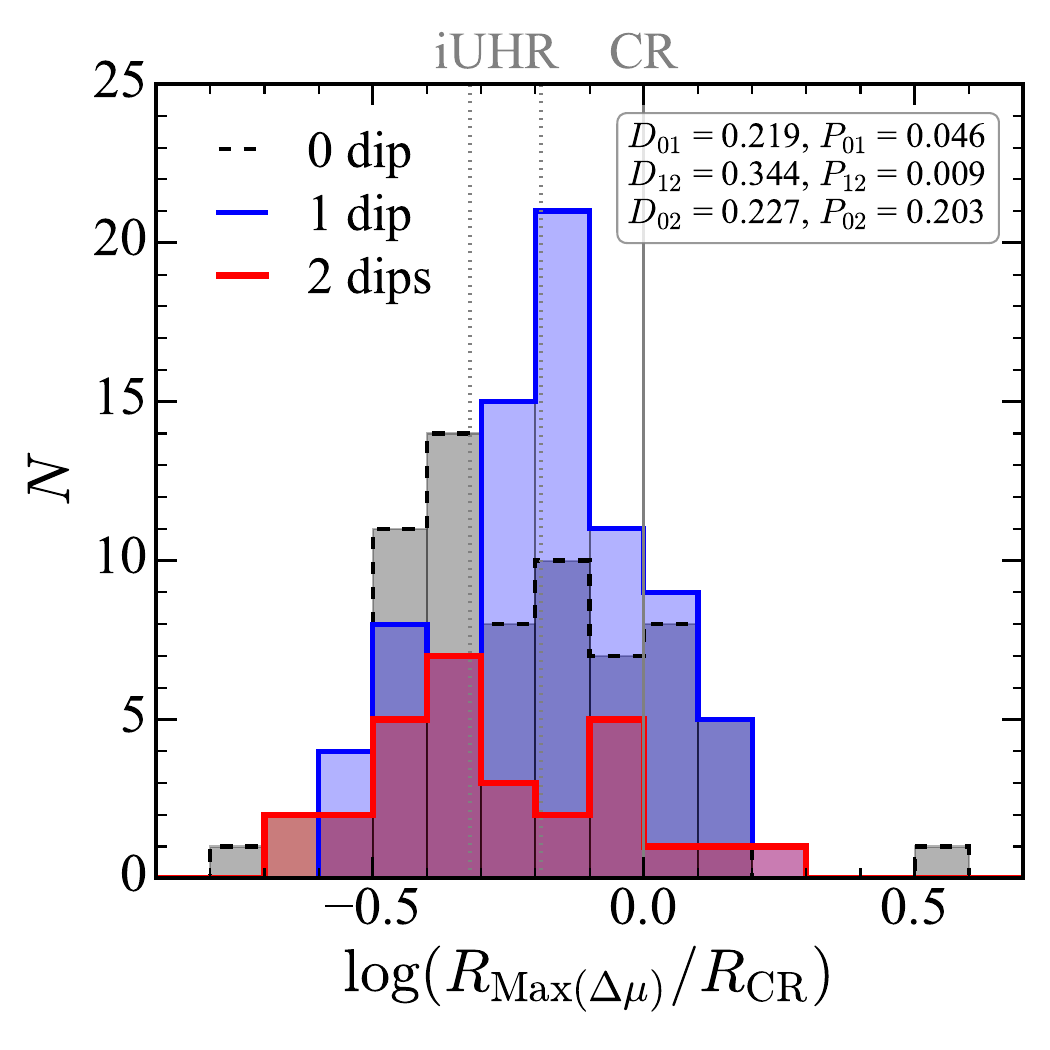}
\caption{The distribution of $\log(R_{\mathrm{Max}(\Delta \mu)}/R_{\mathrm{CR}})$ for galaxies grouped by the number of dips in their surface brightness profiles along the bar minor axis: 0 dip (black dashed line), 1 dip (blue solid line), and 2 dips (red solid line). Vertical dotted lines indicate the region of the inner UHR, and the solid line indicates the CR radius. 
The top-right box shows the results of KS tests between group pairs, including the KS statistic ($D$) and the associated $p$-value ($P$). For example, $D_{01}$ and $P_{01}$ refer to the comparison between the 0-dip and 1-dip groups. The high $D$ value and low $P$ value for $D_{12}$ indicate that the 1-dip and 2-dips groups differ significantly in their distributions.
\label{fig:rmaxmu_rcr_dips}
}
\end{figure}

\begin{figure*}[ht!]
\includegraphics[width=\textwidth]{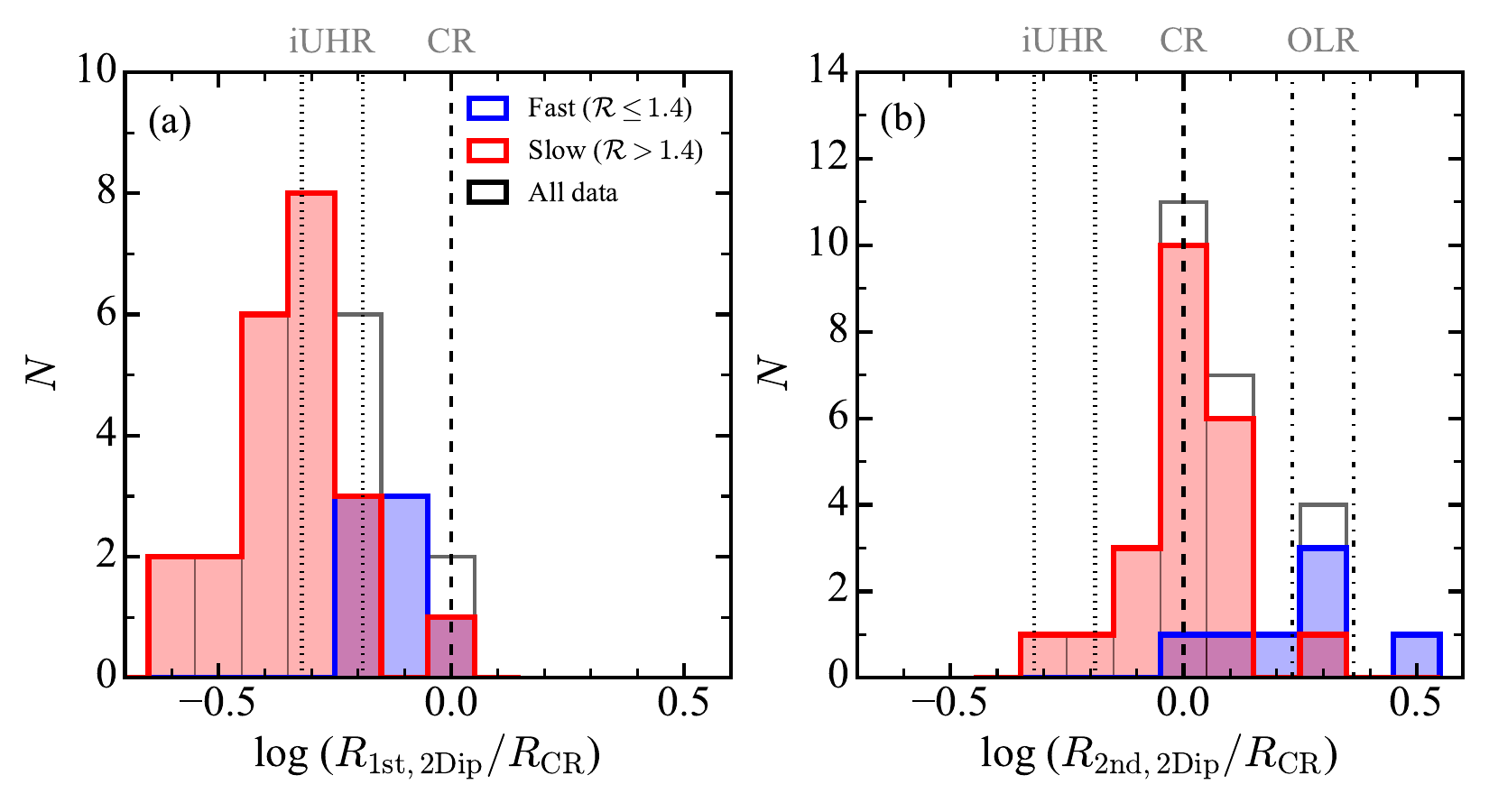}
\caption{Ratio of dark gap radius to corotation radius for galaxies exhibiting two dips. The dip radii are ranked by their distance from the galaxy center.
(a): Histogram of the ratio of the first dip radius to the corotation radius, $R_{\mathrm{1st,2Dip}}/R_{\mathrm{CR}}$.
(b): Histogram of the ratio of the second dip radius to the corotation radius, $R_{\mathrm{2nd,2Dip}}/R_{\mathrm{CR}}$.
The vertical dashed line indicates the corotation radius ($R_{\mathrm{CR}}$), while the dotted and dash-dotted lines correspond to the inner UHR and the outer Lindblad resonance (OLR), respectively.
\label{fig:rdip_rcr_2dips}
}
\end{figure*}

\begin{figure}[hb!]
\includegraphics[width=0.5\textwidth]{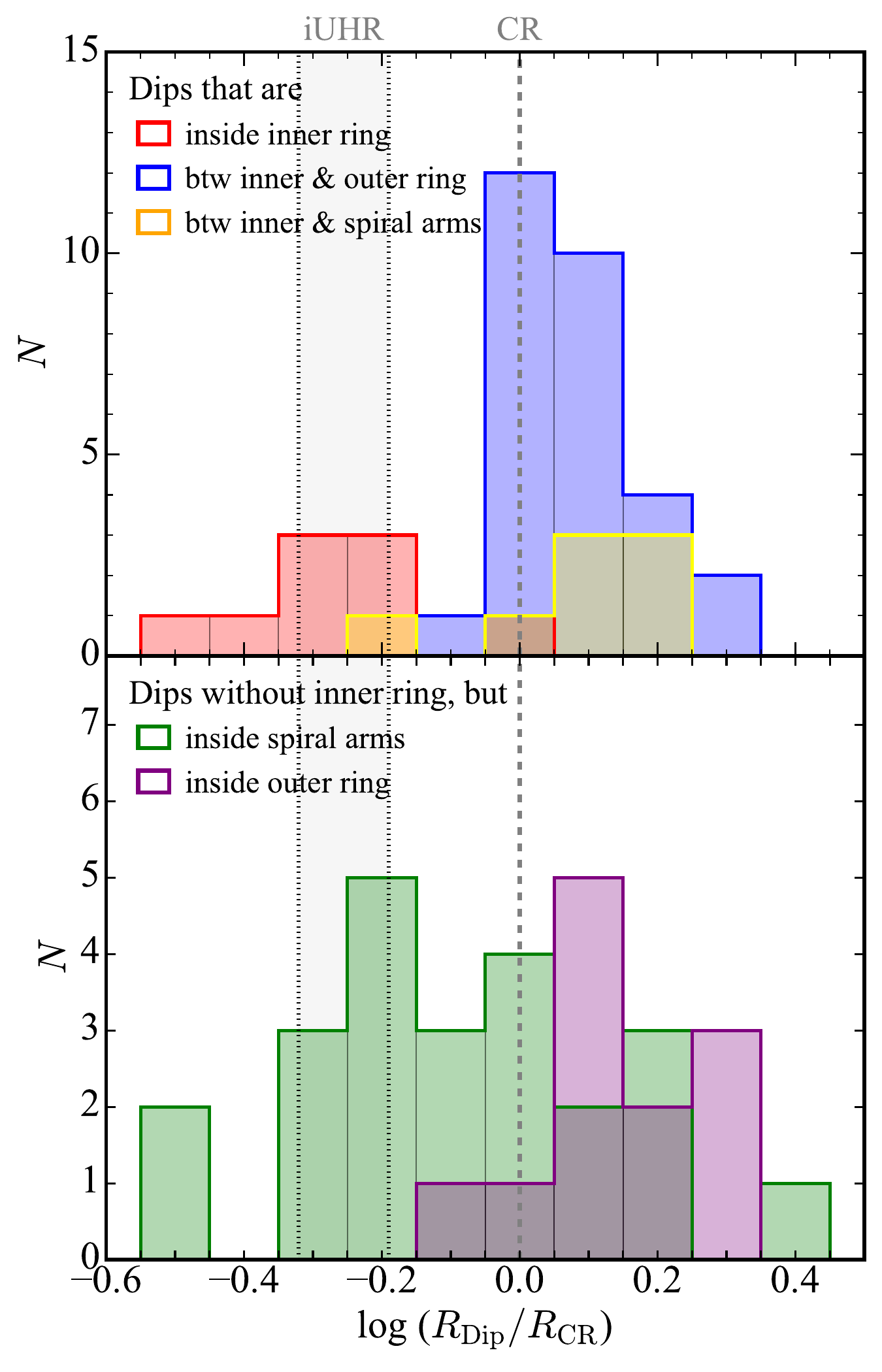}
\caption{
Ratio of $R_{\mathrm{Dip}}/R_{\mathrm{CR}}$ for galaxies exhibiting a single dip. The galaxies are classified into five groups and plotted using different colors. Galaxies with an inner ring are shown in the upper panel, while those without an inner ring are presented in the lower panel. 
\label{fig:rdip_rcr_groups}
}
\end{figure}

\section{Dark gap and resonance radius} 
\label{sec:dg_resonance}

To investigate potential relationships between dark gaps and the dynamical resonances of bars, we compare the radius of the dark gap with resonance radii in this section. We utilize the corotation radius ($R_{\rm CR}$) values reported by \citet{geron_23}. The locations of other dynamical resonances are estimated as follows. The galaxy rotation curve can be approximated by $V \propto R^{-\delta + 1}$, where the exponent $\delta$ characterizes the shape of the radial velocity profile \citep{athanassoula_82}. For a flat rotation curve, $\delta = 1$. Assuming this, the radii of various resonances are derived following the relations given by \citet{athanassoula_82}:

\begin{equation}
\left( \frac{R_{\rm UHR}} {R_{\rm CR}} \right )^ \delta = 1 - \frac{1}{2}\Delta  ~;~~
\left( \frac{R_{\rm OLR}} {R_{\rm CR}} \right ) ^ \delta = 1 + \Delta,
\label{eq:resonances}
\end{equation}

where $\Delta = \left(1 - \frac{1}{2}\delta\right)^{1/2}$.
As shown in Eq.~\ref{eq:resonances}, the locations of dynamical resonances are sensitive to the shape of the rotation curve.
Assuming a power-law rotation curve with $0.7 \leq \delta \leq 1.0$, which is a range typical for Sa to Sb-type galaxies as adopted by \citet{athanassoula_82}, the ratio ${R_{\rm UHR}}/{R_{\rm CR}}$ falls within $0.48 \leq {R_{\rm UHR}}/{R_{\rm CR}} \leq 0.65$.
This is consistent with the representative value ${R_{\rm UHR}}/{R_{\rm CR}} = 0.55$ reported by \citet{krishnarao_22}.
Similarly, for the outer Lindblad resonance (OLR), the estimated range is $1.71 \leq {R_{\rm OLR}}/{R_{\rm CR}} \leq 2.33$.


In Figure~\ref{fig:rdg_rcr}, we compare the corotation radius with the dark gap radius estimated using two different measures: $R_{\rm Max(\Delta \mu)}$ in panel (a), and $R_{\rm Dip}$ in panel (b). All galaxies are included in panels (a) and (c), while only those with a single dip are shown in panels (b) and (d).

When $R_{\rm Max(\Delta \mu)}$ is used as a proxy for the dark gap location, the majority of galaxies have $R_{\rm Max(\Delta \mu)}$ lying between the inner UHR and the CR radius. This shows that $R_{\rm Max(\Delta \mu)}$ is not related to any specific resonance. In Section~\ref{subsec:dg_r}, we will further explore the connection between $R_{\rm Max(\Delta \mu)}$ and the bar length in detail.

Interestingly, when the sample is divided according to their rotation rate ($\mathcal{R}$), slow bars ($\mathcal{R}$ $>$ 1.4) tend to exhibit lower values of $\log(R_{\rm Max(\Delta \mu)}/R_{\rm CR})$, while fast bars ($\mathcal{R} \leq 1.4$) show higher values.

Alternatively, when $R_{\mathrm{Dip}}$ is used as an indicator of the dark gap location, a different trend emerges for galaxies exhibiting a single dip. As shown in Figure~4(b), the distribution displays a strong peak at zero, indicating that $R_{\mathrm{Dip}}$ most commonly coincides with $R_{\mathrm{CR}}$. In addition, a secondary peak is present, suggesting that $R_{\mathrm{Dip}}$ also occurs frequently near the inner UHR. When the sample is further divided according to $\mathcal{R}$, a systematic shift in the distribution is observed between galaxies with higher and lower $\mathcal{R}$.
 
To examine whether the location of dark gaps is related to $\mathcal{R}$, we plot the dark gap radius against $\mathcal{R}$ in Figure~\ref{fig:rdg_rcr}(c) and (d). We find a clear anti-correlation between the two quantities, except for galaxies with extremely slow bars ($\mathcal{R} > 4$). These results suggest that dark gaps do not form at a fixed or universal radius; rather, their location varies systematically with $\mathcal{R}$, the corotation to bar radius. We further investigate the origin of this anti-correlation in Sec.~\ref{subsec:dg_r}.

Figure~\ref{fig:rmaxmu_rcr_dips} shows the distribution of $\log(R_{\mathrm{Max}(\Delta \mu)}/R_{\mathrm{CR}})$ for galaxies categorized into three groups based on the number of dips (0, 1, or 2) identified in their radial profiles along the bar minor axis. A KS test reveals a statistically significant difference between the 1-dip group and the 2-dip group. However, the remaining group pairs involving the 0-dip group (i.e., 0 vs. 1-dip and 0 vs. 2-dip) do not show statistically significant differences, suggesting that the differences in their distributions are not strong enough to be conclusive. This may suggest that galaxies in the 0-dip group are in a transitional phase evolving into systems with 1 or 2 dips, but the bar-driven stellar trapping is not yet sufficiently strong to produce a clearly pronounced dip. 
Alternatively, the evolution may proceed in the opposite direction, i.e., from 1 or 2 dips into none as a result of gas depletion, which suppresses star-forming features particularly at and beyond the bar radius. This may explain why the 0-dip group does not show a statistically significant difference from either the 1-dip or 2-dip group based on the KS test.

For galaxies exhibiting two distinct dips in their radial surface brightness profiles along the bar minor axis, the two dip radii are likely associated with different dynamical resonances. In Figure~\ref{fig:rdip_rcr_2dips}, we present the radii of the first and second dips - ordered by increasing distance from the galaxy center - relative to $R_{\rm CR}$.
We find that among galaxies with two dips, the first dip typically arises inside the (pseudo) inner ring (23 out of 30 cases). In the remaining galaxies, the first dip is located within tightly wound spiral arms.
The distribution of the first dip radius ($R_{\rm 1st,dip}$) peaks near the location of the inner UHR, suggesting a typical placement around the inner UHR. In contrast, the second dip radius ($R_{\rm 2nd,dip}$) is found to coincide with $R_{\rm CR}$ in the majority of cases, indicating a possible link between the second dip and the corotation of the bar.

Although relatively weak, a secondary peak is present near the location of the OLR, primarily contributed by galaxies with fast bars. Consistent with the trend shown in Figure~\ref{fig:rdg_rcr}, galaxies with two dips also tend to have both dip radii located further out in the disk in fast bars compared to slow bars.
In our sample, the fraction of fast bars is relatively low \citep{geron_23}, which may partly account for the weak prominence of the secondary peak.
Nevertheless, it is evident that in galaxies exhibiting two dips, the locations of both the first and second dips vary systematically with the bar rotation rate, $\mathcal{R}$.

We find that $R_{\rm Max(\Delta \mu)}$ does not correspond to any specific resonance when considering all the sample galaxies in Section~\ref{sec:dg_resonance}. Although $R_{\rm Max(\Delta \mu)}$ lies between the inner UHR and CR, this may be coincidental, as both resonances and dark gaps are confined to the bar region. Thus, some spatial proximity is naturally expected. In the case of $R_{\rm Max(\Delta \mu)}$, this proximity may reflect a random statistical distribution rather than a direct physical association. This might also apply, to some extent, to the dark radii defined by the location of the dip (e.g., $R_{\rm dip}$, $R_{\rm 1st,dip}$, $R_{\rm 2nd,dip}$). However, in contrast to $R_{\rm Max(\Delta \mu)}$, we find relatively well-defined clustering of $R_{\rm dip}$ with CR, $R_{\rm 1st,dip}$ with the inner UHR, and $R_{\rm 2nd,dip}$ with CR.

\section{Dark gap and the galaxy morphology}

We explore the influence of morphology of barred galaxies on the association between $R_{\rm dip}$ and dynamical resonances by assessing whether different bar types exhibit consistent trends in dip positions.

To this end, we classify galaxies with a single dip into five categories based on the position of the dip :
1) located inside the inner ring (the ring at the ends of the bar),
2) between the inner ring and the (pseudo) outer ring,
3) between the inner ring and the spiral arms,
4) within the outer ring in galaxies without an inner ring, and
5) within the spiral arms in galaxies without an inner ring.

Figure~\ref{fig:rdip_rcr_groups} presents the normalized dip radius ($R_{\rm Dip}/R_{\rm CR}$) for galaxies exhibiting a single dip, with the upper panel showing those with an inner ring and the lower panel showing those without one.
Galaxies with a dip located inside the inner ring (shown in red in Figure~\ref{fig:rdip_rcr_groups}) tend to have $R_{\rm Dip}$ near the inner UHR. Interestingly, galaxies with a dip located between the inner and outer rings (blue) predominantly show $R_{\rm Dip}$ at $R_{\rm CR}$ or slightly beyond it. For galaxies where the dip lies between the inner ring and the spiral arms (yellow), $R_{\rm Dip}$ is generally found beyond $R_{\rm CR}$. In contrast, galaxies lacking an inner ring and exhibiting a dip within the spiral arms (green) show no clear preference in the dip location; the $R_{\rm Dip}/R_{\rm CR}$ values span a broad range. Meanwhile, galaxies without an inner ring but with a dip inside the outer ring (purple) tend to have $R_{\rm Dip}$ located beyond $R_{\rm CR}$.

Distinct ranges of $R_{\rm Dip}/R_{\rm CR}$ are found for most morphological groups, with the exception of galaxies exhibiting a dip within the spiral arms in the absence of an inner ring, which display a broad dispersion in the dip location.
Our findings suggest that the relationship between $R_{\rm Dip}$ and $R_{\rm CR}$ depends on specific morphological features of the galaxy, particularly the presence and configuration of bars and rings (inner and outer) if present, and the relative position of the dark gap with respect to these structures.
Thus, morphological information may allow us to approximate estimates of the CR radius for certain types of galaxies. 
The resonance radii change as the galaxy evolves. Resonance locations are modified in response to variations in the gravitational potential, which result from angular momentum transfer driven by the bar and the evolving mass distribution of the galaxy (e.g., \citealt{dubinski_09}). Therefore, the resonance radius would be modified accordingly as the galaxy evolves, and the amount of change would differ from galaxy to galaxy, depending on the properties of individual galaxies, including morphology.
As we have seen that the dip radius to the corotation radius is associated with $\mathcal{R}$ in Figure \ref{fig:rdg_rcr}, our results suggest that galaxies with similar morphological characteristics, specifically in terms of ring structure, tend to exhibit similar values of $\mathcal{R}$, which in turn results in comparable dark gap radii.
This morphological dependence may provide a basis for future studies aimed at improving our understanding of the underlying dynamical processes that shape barred galaxies and refining models of resonance-driven structures.

\section{Discussion}
\subsection{Correlation Between Bar Length and Dark Gap Radius
}
\label{subsec:dg_r}
\begin{figure*}[ht!]
\includegraphics[width=\textwidth]{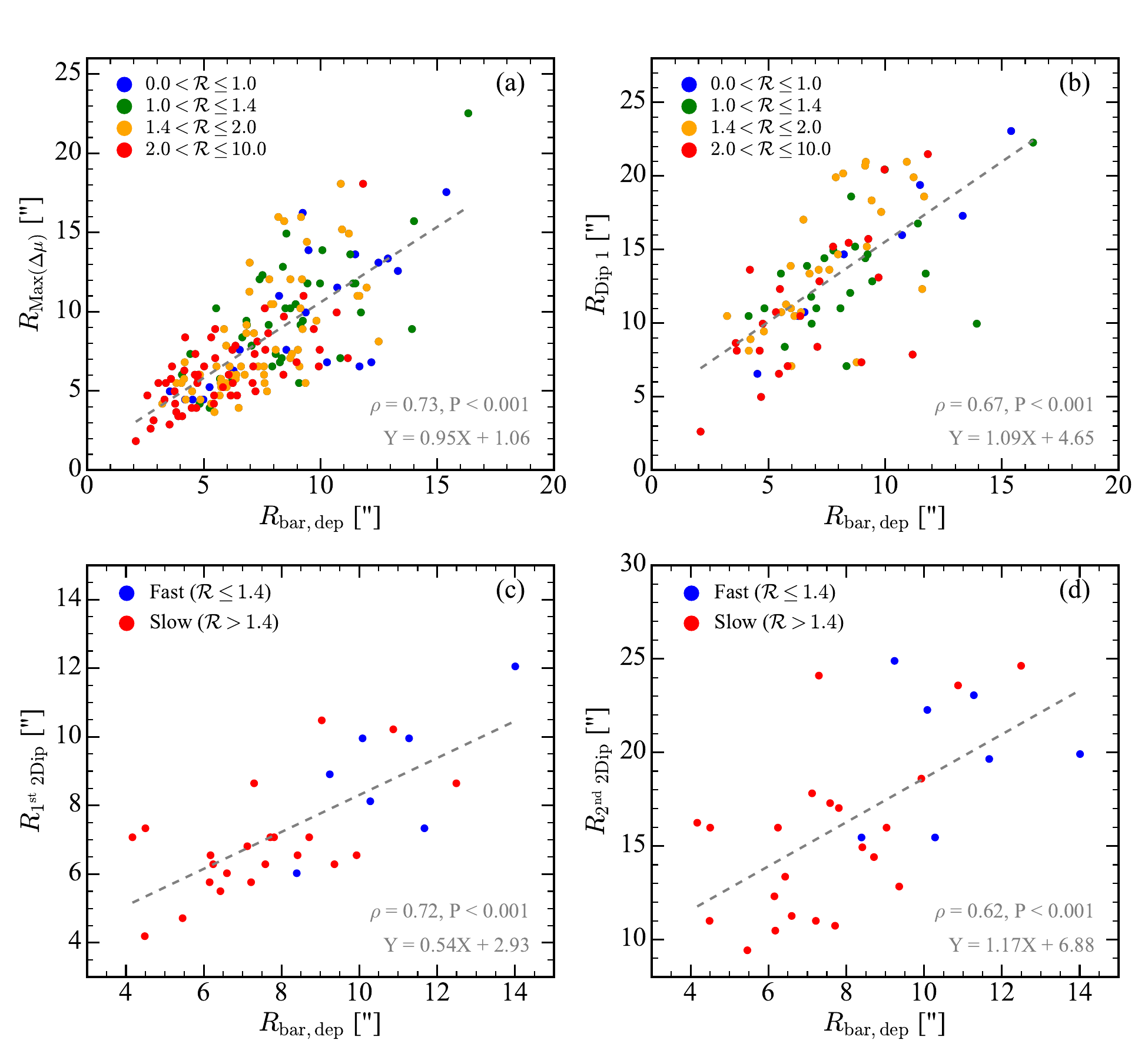}
\caption{Bar radius and dark gap radius which is traced with
(a): $R_{\rm Max (\mu)}$ for all galaxies,
(b): $R_{\rm Dip}$ for galaxies exhibiting a single dip along the bar minor axis,
(c): the first dip position, $R_{\rm 1^{st} Dip 2}$,, for galaxies with two dips, and
(d): the second dip, $R_{\rm 2^{nd} Dip 2}$, for galaxies with two dips.  Pearson correlation coefficients and corresponding linear fits are shown in the lower right corner of each panel.}
\label{fig:fig7_dg_rbar}
\end{figure*}

In Section~\ref{sec:dg_resonance}, we find that the location of the dark gap varies with $\mathcal{R}$, as illustrated in Figure~\ref{fig:rdg_rcr}, showing that slow bars tend to exhibit lower values of $R_{\rm Max(\Delta \mu)}/R_{\rm CR}$ compared to fast bars.
In particular, we find that $\log(R_{\rm Max(\Delta \mu)}/R_{\rm CR})$ decreases linearly with increasing $\mathcal{R}$.
As $\mathcal{R}$ is defined as $R_{\rm CR}/R_{\rm bar}$, this anti-correlation may reflect an underlying relation between the bar radius and the location of the dark gap. To test this possibility, we plot the deprojected bar radius against the dark gap radius estimated by two methods in Figure~\ref{fig:fig7_dg_rbar}.
The dark gap radii, characterized by both $R_{\rm Max(\Delta \mu)}$ and $R_{\rm Dip}$, show a clear correlation with the bar radius.
This trend also holds for galaxies exhibiting two dips, with both the first and second dip radii following the same general relation, albeit with larger scatter. Figure~\ref{fig:fig7_dg_rbar}(d) shows that the second dip radius is well beyond the bar radius.

Pearson correlation analysis reveals a strong and statistically significant relation (P $<$ 0.001) between the bar radius and the dark gap radius, indicating that galaxies with longer bars tend to host dark gaps at larger radii.
This correlation has also been investigated in numerical simulations \citep{kim_16, ghosh_24b}.
\citet{ghosh_24b} report that the relation generally holds, except in extreme cases where the thick disk contributes up to 90$\%$ of the total disk mass which is considered exceptionally high for disk galaxies \citep{yoachim_06, comeron_14b}.
Consistent with these findings, our observational analysis across a diverse sample of barred galaxies confirms a tight correlation between bar length and dark gap radius. 
This relation also supports the hypothesis that dark gaps arise from bar-driven dynamics.

\subsection{Discrepancies arising from heterogeneous samples}

Several studies have investigated the locations of dark gaps and their relation to dynamical resonances. However, observational and simulation-based studies have yielded seemingly contradictory interpretations, as briefly outlined in the Introduction.
\citet{buta_17b} argues that dark gaps can be used to infer the corotation radius of the bar in certain galaxies, particularly in cases where a dark gap is present between the inner and outer rings.
In that study, the dark gap radius was identified by locating the minimum in surface brightness along the bar minor axis using a parabolic fit, a method that corresponds to our definition of $R_{\rm Dip}$.
The interpretation by \citet{buta_17b} that dark gaps trace the corotation radius is consistent with the subset of our sample where dips are found between the inner and outer rings, as shown by the blue line in Figure~\ref{fig:rdip_rcr_groups}.

However, \citet{ghosh_24b} report from their numerical simulations that none of the bar-associated resonances coincide with the locations of dark gaps. Their analysis specifically focuses on dark gaps identified using $R_{\rm Max(\Delta \mu)}$ in galaxies with slow bars. Consistent with this, when we restrict our sample to slow bars in Figure~\ref{fig:rdg_rcr}(a), we also find no clear preference for the location of the dark gap traced by $R_{\rm Max(\Delta \mu)}$.

An alternative interpretation regarding the location of dark gaps has been proposed by \citet{krishnarao_22} and \citet{aguerri_23}, who argue that the majority of dark gaps are closely associated with the UHR of the bar. The numerical simulations analyzed by \citet{krishnarao_22} do not develop a ring structure, which is different from the assumption in the work of \citet{buta_17b}. Consequently, this difference may result in alternative interpretations of the connection between dark gaps and resonances.
Because \citet{krishnarao_22} analyze their observational sample galaxies as an ensemble, it is challenging to identify the specific characteristics of their sample in terms of the rotation parameter $\mathcal{R}$ or ring morphology. Similarly, as the sample characterization in \citet{aguerri_23} is not detailed, we can only speculate that their galaxies may cover limited or specific ranges of $\mathcal{R}$.

Although these studies seem to yield different results and interpretations, the discrepancies mainly stem from differences in sample selection and the definition of dark gap radii. Therefore, the apparent contradictions do not reflect fundamental disagreements but rather arise from variations in the observational scope and sample characteristics.

\subsection{Limitation}
Several factors may contribute to uncertainties in determining the locations of dark gaps and resonances.

First, we use the ellipticity of galaxy isophotes to deproject the images, under the assumption that the disk is intrinsically circular. However, the measured ellipticity may be affected by structural features such as spiral arms and outer rings. In particular, outer rings often exhibit dimples ($R_1$ ring, \citealt{buta_15, buta_17b}), which cause the outer isophotes to appear more elongated intrinsically. Consequently, this can introduce uncertainties in the derived locations of dark gaps. 
These factors can introduce systematic errors into the TW method with the uncertainty in the position angle of the line of nodes additionally contributing to inaccuracies in $\Omega_{\rm bar}$ \citep{zou_19} and the derived $R_{\rm CR}$.

Second, the epicyclic approximation used to estimate the resonance radii is valid only under the assumption of small perturbations (\citealt{binney_87}). Consequently, applying this approximation to systems with strong bars may result in uncertainties in estimates of resonance locations (\citealt{ruiz-garcia_24}).
We also noet that even in cases where the resonance condition is only approximately satisfied, “near-resonance effects” \citep{contopoulos_80a} can arise, exhibiting phenomena similar to those under exact resonance, including phase-space splitting, orbit trapping, and irregular orbital behavior.

Third, resonant orbits often deviate from perfectly circular shapes. For example, ILR orbits tend to be elongated (e.g., \citealt{maciejewski_02, li_15, struck_15}), while orbits associated with CR, UHR, and OLR form broad rings that occupy a finite radial extent (e.g., \citealt{ceverino_07}). This complicates the assignment of a unique radius to each resonance, as the resonant regions themselves span a range of radii rather than a single, well-defined location.



\subsection{Why not all barred galaxies show dark gaps?}
\label{subsec:no_dg}
Although we find that the majority of barred galaxies ($61\%$) exhibit at least one dark gap, it is evident that this feature is not present in all barred galaxies. In our analysis, galaxies that show a monotonically decreasing surface brightness profile along the bar minor axis, without a clearly identifiable dip, are classified as lacking a dark gap. 
However, we note that although these galaxies do not display a clear dip in the radial profile along the minor axis, they still exhibit a noticeable light deficit compared to the profile along the bar major axis.
Galaxies without a dark gap show no clear preference in bar strength: 45$\%$ (34/75) are classified as strong bars and 55$\%$ (41/75) as weak bars, based on the classification by \citet{geron_23}.
Since dark gaps are located relatively close to the inner part of the disk and our data extend deep enough to cover the outer ring, the absence of dark gaps in some barred galaxies is unlikely to be due to limitations in observational depth.

This can be attributed to several factors, which are outlined below.
First, as galaxies evolve, bars can trap nearby stars into their orbits. As a result, the more stars that are captured from the inner disk, the more prominent the resulting dark gaps tend to be. Therefore, galaxies that do not exhibit a clear dark gap may be at an earlier stage of bar evolution and may not have had sufficient time to evacuate stars from the inner disk regions surrounding the bar.

Second, an additional mechanism that may influence the radial distribution of stars is radial migration, a process by which stars shift their orbital radii over time (e.g., \citealt{sellwood_02}).
Stars can travel significant radial distances across galactic disks as a result of migration driven by resonant scattering with spiral arms (\citealt{roskar_08b, grand_12}) or through the overlapping of spiral and bar resonances (\citealt{minchev_10}). If stellar migration is strong, faint dark gaps may be filled in by stars originating from other parts of the disk, particularly in galaxies where the bar has not yet efficiently captured stars. 

Third, in the absence of an outer ring or well-defined spiral arms, there may be no corresponding enhancement in stellar density in the radial surface brightness profile at those locations. Consequently, the lack of such structures may simply result in the absence of a detectable dark gap. 
Therefore, we cannot rule out the possibility that the mere presence of outer rings and spiral arms may contribute to the appearance of dark gaps. However, the left two panels of Figure~\ref{fig:three_in_row} clearly show a distinct difference in the radial profile along the bar minor axis. The position of the dip is located farther out, and the surface brightness at the dip is relatively low, clearly indicating a depletion of stars. Thus, the presence of an outer ring or spiral arms alone cannot reproduce the features observed in galaxies with a single dip. Further numerical simulations are necessary to better understand the origin of galaxies lacking such dips.

\section{Summary and Conclusion}
Dark gaps are found in the majority of barred galaxies. Previous studies have proposed a link between the radius of dark gaps and bar resonances, although their conclusions have varied. We compare the CR radius of the bar derived via the TW method applied to the MaNGA dataset, with the dark gap radii obtained from the DESI Legacy Imaging Survey. Our main results are summarized below.

\begin{enumerate}
\item Dark gap radii can be estimated using two distinct methods: (i) the radius at which the surface brightness difference between the bar major and minor axes reaches its maximum, ($R_{\rm Max(\Delta \mu)}$), and (ii) the radius corresponding to a local minimum (dip) in the bar minor axis profile ($R_{\rm Dip}$). For the majority of galaxies in our sample, the two estimates differ by more than 0.3$\times R_{\rm bar}$, indicating that their application and interpretation require careful consideration.

\item Our results indicate that the positions of dark gaps are not governed by any specific dynamical resonance. Instead, the dark gap radii are correlated with the rotation rate of the bar, $\mathcal{R}$, which represents the ratio of the corotation radius to the bar length. In galaxies with slow bars, the dark gaps are typically located well inside the corotation radius of the bar, whereas in those with fast bars, they appear much closer to it. This trend suggests that dark gap locations are fundamentally linked to bar length, rather than to fixed resonance radii.

\item Classification of galaxies based on ring morphology reveals that specific morphological types tend to exhibit dark gaps aligned with particular dynamical resonances. For example, when a dark gap is located between the inner and outer $R_1$ rings, it is closely associated with the corotation radius. In galaxies showing two dips in the bar minor axis profile, the first dip typically corresponds to the UHR, while the second aligns with the CR. These results indicate that certain morphological types share similar values of $\mathcal{R}$, and consequently, the dark gap radius in such systems is strongly linked to bar-driven dynamical resonances.

\item Previous studies have reported conflicting interpretations regarding the location of dark gaps and their association with resonances. However, these discrepancies arise primarily from differences in sample selection, definitions of dark gap radii, and morphological coverage, rather than from fundamental inconsistencies in the underlying physics. 
Our results help reconcile some of these differences by demonstrating that the resonance associated with a dark gap depends on both the identification method and the ring morphology of barred galaxies.

\end{enumerate}

\begin{acknowledgments}
This research was supported by Kyungpook National University Development Project Research Fund, 2024.
The imaging data used in this study are publicly available from the DESI Legacy Imaging Surveys at \url{https://www.legacysurvey.org/}. 

\end{acknowledgments}

%
\facilities{Sloan, Mayall, Bok, Blanco}

\software{Source Extractor \citep{bertin_96}, Astropy \citep{astropy_col_13, astropy_col_18,astropy_col_22}, Scipy \citep{Virtanen_20}, Matplotlib \citep{hunter_07}, Numpy \citep{vanderwalt_11}
          }




\bibliography{tkim_bar25}{}
\bibliographystyle{aasjournalv7}



\end{document}